\documentclass[useAMS,usenatbib]{mn2e}
\usepackage{amssymb}
\usepackage{graphicx}
\usepackage{amsmath}
\usepackage{color}
\usepackage{cancel}
\usepackage[normalem]{ulem}

\def\spose#1{\hbox to 0pt{#1\hss}}
\def\ltwig{\mathrel{\spose{\lower 3pt\hbox{$\mathchar"218$}}
     \raise 2.0pt\hbox{$\mathchar"13C$}}}
\def\gtwig{\mathrel{\spose{\lower 3pt\hbox{$\mathchar"218$}}
     \raise 2.0pt\hbox{$\mathchar"13E$}}}
\def\q{p}
\def\cq{C_{\rm p}}
\def\Rstar{R_{\ast}}
\def\Mstar{M_{\ast}}

\def\Mdot{\dot M}

\def\solar{\odot}
\def\Msun{M_{\solar}}

\def\Rsun{R_{\solar}}

\def\vinf{V_\infty}

\newcommand{\beq}{\begin{equation}}
\newcommand{\eeq}{\end{equation}}
\newcommand{\beqa}{\begin{align}}
\newcommand{\eeqa}{\end{align}}

\title[Thin-shell mixing and O-star X-rays]{
Thin-shell mixing in radiative wind-shocks and the $L_{\rm x} \sim L_{\rm bol}$ scaling of O-star X-rays
}
\author[Owocki et al.]{S.\ P.\ Owocki$^1$, J.\  O.\  Sundqvist$^1$,  D.\ H.\ Cohen$^2$,
and K.\ G.\ Gayley$^3$
\\
$^1$Bartol Research Insitute, 
Department of Physics \& Astronomy, 
University of Delaware, Newark,DE 19716 USA
\\
$^2$Department of Physics \& Astronomy, Swarthmore College, Swarthmore, PA 19081 USA
\\
$^3$Department of Physics, University of Iowa, Iowa City, IA
52242 USA
}

\begin{document}

%
%
%


\def\jnl@style{\it}
\def\aaref@jnl#1{{\jnl@style#1}}

\def\aaref@jnl#1{{\jnl@style#1}}

\def\aj{\aaref@jnl{AJ}}                   
\def\araa{\aaref@jnl{ARA\&A}}             
\def\apj{\aaref@jnl{ApJ}}                 
\def\apjl{\aaref@jnl{ApJ}}                
\def\apjs{\aaref@jnl{ApJS}}               
\def\ao{\aaref@jnl{Appl.~Opt.}}           
\def\apss{\aaref@jnl{Ap\&SS}}             
\def\aap{\aaref@jnl{A\&A}}                
\def\aapr{\aaref@jnl{A\&A~Rev.}}          
\def\aaps{\aaref@jnl{A\&AS}}              
\def\azh{\aaref@jnl{AZh}}                 
\def\baas{\aaref@jnl{BAAS}}               
\def\jrasc{\aaref@jnl{JRASC}}             
\def\memras{\aaref@jnl{MmRAS}}            
\def\mnras{\aaref@jnl{MNRAS}}             
\def\pra{\aaref@jnl{Phys.~Rev.~A}}        
\def\prb{\aaref@jnl{Phys.~Rev.~B}}        
\def\prc{\aaref@jnl{Phys.~Rev.~C}}        
\def\prd{\aaref@jnl{Phys.~Rev.~D}}        
\def\pre{\aaref@jnl{Phys.~Rev.~E}}        
\def\prl{\aaref@jnl{Phys.~Rev.~Lett.}}    
\def\pasp{\aaref@jnl{PASP}}               
\def\pasj{\aaref@jnl{PASJ}}               
\def\qjras{\aaref@jnl{QJRAS}}             
\def\skytel{\aaref@jnl{S\&T}}             
\def\solphys{\aaref@jnl{Sol.~Phys.}}      
\def\sovast{\aaref@jnl{Soviet~Ast.}}      
\def\ssr{\aaref@jnl{Space~Sci.~Rev.}}     
\def\zap{\aaref@jnl{ZAp}}                 
\def\nat{\aaref@jnl{Nature}}              
\def\iaucirc{\aaref@jnl{IAU~Circ.}}       
\def\aplett{\aaref@jnl{Astrophys.~Lett.}} 
\def\apspr{\aaref@jnl{Astrophys.~Space~Phys.~Res.}}
\def\bain{\aaref@jnl{Bull.~Astron.~Inst.~Netherlands}} 
\def\fcp{\aaref@jnl{Fund.~Cosmic~Phys.}}  
\def\gca{\aaref@jnl{Geochim.~Cosmochim.~Acta}}   
\def\grl{\aaref@jnl{Geophys.~Res.~Lett.}} 
\def\jcp{\aaref@jnl{J.~Chem.~Phys.}}      
\def\jgr{\aaref@jnl{J.~Geophys.~Res.}}    
\def\jqsrt{\aaref@jnl{J.~Quant.~Spec.~Radiat.~Transf.}}
\def\memsai{\aaref@jnl{Mem.~Soc.~Astron.~Italiana}}
\def\nphysa{\aaref@jnl{Nucl.~Phys.~A}}   
\def\physrep{\aaref@jnl{Phys.~Rep.}}   
\def\physscr{\aaref@jnl{Phys.~Scr}}   
\def\planss{\aaref@jnl{Planet.~Space~Sci.}}   
\def\procspie{\aaref@jnl{Proc.~SPIE}}   

\let\astap=\aap
\let\apjlett=\apjl
\let\apjsupp=\apjs
\let\applopt=\ao

\date{Accepted ?.  Received ?; in original form ?}

\pagerange{\pageref{firstpage}--\pageref{lastpage}} \pubyear{2011}

\maketitle

\label{firstpage}

\begin{abstract}

X-ray satellites since {\em Einstein} have empirically established that the X-ray luminosity from single O-stars scales linearly with bolometric luminosity, $L_{\rm x} \sim 10^{-7} L_{\rm bol}$.  But straightforward forms of the most favored model, in which X-rays arise from instability-generated shocks embedded in the stellar wind, predict a steeper scaling, either with mass loss rate $L_{\rm x} \sim \Mdot \sim L_{\rm bol}^{1.7}$ if the shocks are radiative, or with $L_{\rm x} \sim \Mdot^{2} \sim  L_{\rm bol}^{3.4}$ if they are adiabatic. This paper presents a generalized formalism that bridges these radiative vs.\ adiabatic limits in terms of the ratio of the shock cooling length to the local radius.
Noting that the thin-shell instability of radiative shocks should lead to extensive mixing of hot and cool material, we propose that the associated softening and weakening of the X-ray emission can be parameterized as scaling with the cooling length ratio raised to a power $m$, the ``mixing exponent."   For physically reasonable values $m \approx 0.4$, this leads to  an X-ray luminosity $L_{\rm x} \sim \Mdot^{0.6} \sim L_{\rm bol}$ that matches the empirical scaling.  To fit observed X-ray line profiles, we find such radiative-shock-mixing models require the number of shocks to drop sharply above the initial shock onset radius. This in turn implies that the X-ray luminosity should saturate and even decrease for optically thick winds with very high mass-loss rates. In the opposite limit of adiabatic shocks in low-density winds (e.g., from B-stars), the X-ray luminosity should drop steeply with $\Mdot^2$. Future numerical simulation studies will be needed to test the general thin-shell mixing
{\em ansatz} for X-ray emission.
\end{abstract}


\section{Introduction}

Since the 1970's X-ray satellite missions like {\it Einstein}, {\it ROSAT}, and most recently {\it Chandra} and {\it XMM-Newton} have found hot, luminous, O-type stars to be sources of soft ($\ltwig 1$~keV) X-rays, with a  roughly\footnote{i.e., extending over 2 dex in $L_{\rm bol}$, with a typical scatter of $\sim \pm 0.5$~dex} {\em linear} scaling between the X-ray luminosity and the stellar bolometric luminosity, $L_{\rm x} \sim 10^{-7} L_{\rm bol}$ 
 \citep{Chlebowski89,  Kudritzki06, Berghoefer97, Sana06, Gudel09, Naze09, Naze11}.
In some systems with harder  (a few keV) spectra and/or higher $L_ x$, the observed X-rays have been associated with shock emission in colliding wind binary (CWB) systems 
\citep{Stevens92, Gagne11},
 or with magnetically confined wind shocks (MCWS)  
 \citep{Babel97, Wade11}.
But in putatively single, non-magnetic O-stars, the most favored model is that the X-rays are emitted from {\em embedded wind shocks}  that form from the strong, intrinsic instability (the ``line-deshadowing instability" or LDI)  associated with  the driving of these winds by line-scattering of the star's radiative flux 
 \citep{Owocki88, Feldmeier97, Dessart03, Sundqvist12}.

\begin{figure*}
\begin{center}
\includegraphics[scale=0.66]{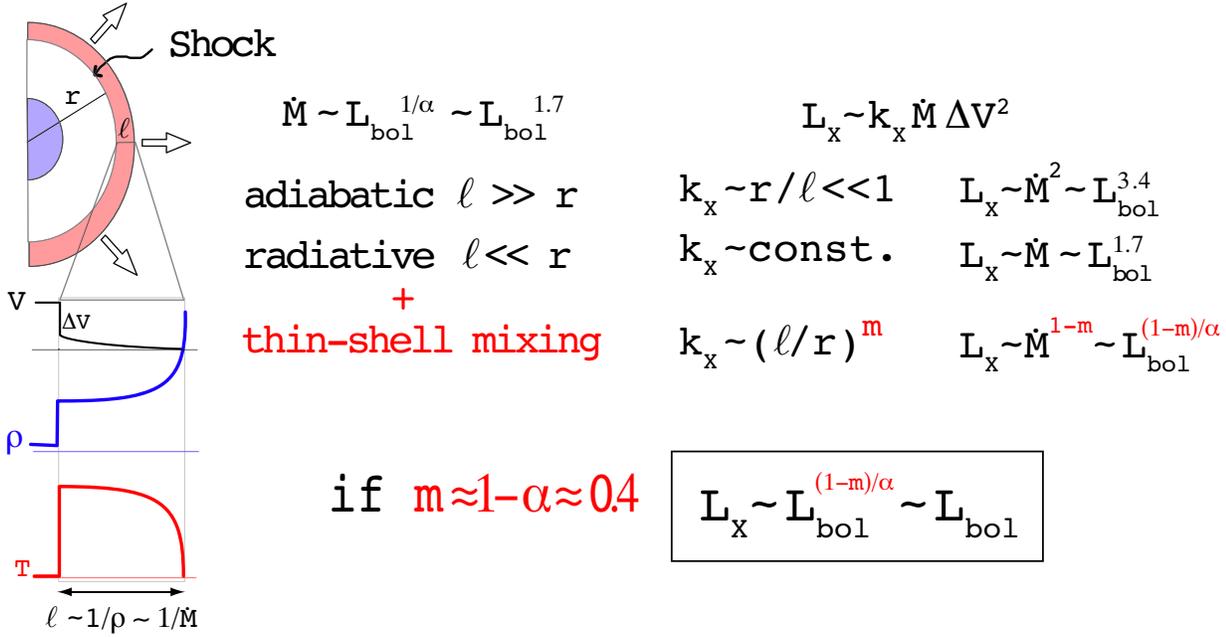}
\caption{
Summary sketch of the key concepts and results of the scaling analysis in this paper. 
The illustration of the cooling zone from a wind shock shows associated scalings for X-ray luminosity $L_{\rm x}$ with mass loss rate ${\dot M}$ and bolometric luminosity $L_{\rm bol}$, based on the conversion factor $k_x$ of wind kinetic energy into X-rays, which for radiative shocks is a constant, but for adiabatic shocks is reduced by the ratio of the radius to cooling length, $r/\ell \ll 1$. (See eqn. \ref{kxdef} in  text.)
In addition,  thin-shell mixing of such radiative shocks is then posited to lead to a reduction of the X-ray emitting fraction that scales as a power-law of  the cooling length, $k_{\rm x} \sim \ell^{m}$.
For CAK wind index $\alpha$, a mixing exponent $m = 1-\alpha$ leads to the observationally inferred linear scaling of X-rays with bolometric luminosity, $L_{\rm x} \sim L_{\rm bol}$.
}
\label{lxlbol-tsm}
\end{center}
\end{figure*}

This LDI can be simply viewed as causing some small ($\ltwig 10^{-3}$) fraction of the wind material to pass through an X-ray emitting shock, implying in the case that the full shock energy is suddenly radiated away that the X-ray luminosity should scale with the wind mass loss rate, $L_{\rm x} \sim \Mdot$.
But within the standard 
\citet[hereafter CAK]{Castor75}
model for such radiatively driven stellar winds, this mass loss rate increases with luminosity\footnote{For simplicity,  this ignores a secondary scaling of luminosity with mass; see \S \ref{md2lbol}.} as $\Mdot \sim L_{\rm bol}^{1/\alpha} \sim L_{\rm bol}^{1.7}$, where the latter scaling uses a typical CAK power index $\alpha \approx 0.6$
\citep{Puls00}.
This then implies a {\em super-linear} scaling for X-ray to bolometric luminosity, $L_{\rm x} \sim L_{\rm bol}^{1.7}$, which is too steep to match the observed, roughly linear law.

In fact, the above sudden-emission scaling 
effectively assumes the shocks are {\em radiative}, with a cooling length  that is much smaller than the local radius, $ \ell \ll r$.
In the opposite limit  $\ell \gg r$, applicable to lower-density winds for which shocks  cool by {\em adiabatic} expansion, the shock emission scales with the X-ray {\em emission measure}, 
EM$\, \sim \int \rho^2 dV$, leading then to an even steeper scaling of X-ray vs. bolometric luminosity, $L_{\rm x} \sim \Mdot ^2 \sim L_{\rm bol}^{3.4}$.

Both these scalings ignore  the effect of bound-free absorption of X-rays by the cool, unshocked material that represents the bulk of the stellar wind.
\citet[hereafter OC99]{Owocki99b}
 showed that accounting for wind absorption can lead to an observed X-ray luminosity that scales linearly  with $L_{\rm bol}$,
 but this requires specifying {\em ad hoc} a fixed radial decline in the volume filling factor for X-ray emitting gas.
 We show below (\S \ref{sec:etaxss}) that this filling factor should actually be strongly affected by 
the level of  radiative cooling.
Moreover, while modern observations of spectrally resolved X-ray emission profiles by {\it Chandra} and {\it XMM-Newton} do indeed show the expected broadening from shocks embedded in the expanding stellar wind, the relatively weak blue-red asymmetry indicates that  absorption effects are modest in even the densest OB-star winds
\citep{Cohen10, Cohen11}.
Since many stars following the $L_{\rm x}$-$L_{\rm bol}$ empirical law have weaker winds that are largely optically thin to X-rays
\citep{Cohen97, Naze11}, it now seems clear that absorption cannot explain this broad $L_{\rm x}$ scaling.

The analysis here examines instead the role of radiative cooling, and associated thin-shell instabilities \citep{Vishniac94, Walder98, Schure09, Parkin10}, 
in mixing shock-heated material with cooler gas, and thereby reducing and softening the overall X-ray emission. As summarized in figure 1, for a simple parameterization that this mixing reduction scales with a power (the ``mixing exponent'' $m$) of the cooling length, $\ell^m$, we find that the linear  $L_{\rm x}$-$L_{\rm bol}$ law can be reproduced by assuming $m \approx 0.4$.
The development below quantifies and extends a preliminary conference presentation of this thin-shell mixing {\em ansatz} \citep{Owocki11}.

Specifically, to provide a quantitative basis for bridging the transition between radiative and adiabatic shock cooling, 
the next section (\S 2) first analyzes in detail the X-ray emission from a simplified model of a single, standing shock in steady, spherically expanding outflow.
The following section (\S 3) then generalizes the resulting simple bridging law to account for thin-shell mixing, and applies this 
in a simple exospheric model to derive general scalings for X-ray luminosity from a wind with multiple, instability-generated shocks assumed to have a power-law number distribution in wind radius.
A further application to computation of X-ray line profiles  (\S 4) provides constraints on the mixing and shock-number exponents needed to match observed X-ray emission lines.
Following a brief presentation (\S 5) of full integral solutions for X-ray luminosity to complement the general scaling laws in \S 3,
the final section (\S 6) concludes with a brief summary and outlook for future work.

\section{Standing Shock Model}

\subsection{Energy Balance}

Most previous analyses of X-rays from massive stars
\citep[e.g., ][]{Wojdowski05, Cohen11} have been cast in terms of a density-squared emission measure from some fixed volume filling factor for X-ray emitting gas 
\citep[see however][]{Krolik85, Hillier93, Feldmeier97b, Antokhin04}.
So let us begin by showing explicitly how that picture must be modified to account for the effects of radiative cooling, which can be important and even dominant for O-star wind shocks\footnote{This was actually noted explicitly by \citet{Zhekov07}, but their results were nonetheless still cast in terms of a density-squared emission measure that is not appropriate for radiative shocks.}.
By focusing on the simple example of a steady, {\em standing} wind-shock, it is possible to carry out an analytic analysis that derives, more or less from first principles, a simple bridging law between the scalings for adiabatic vs.\ radiative shocks.

Specifically, let us consider an idealized model in which a spherically symmetric, steady-state stellar wind with mass loss rate  $\Mdot$ and constant, highly supersonic flow speed $\vinf$ undergoes a strong, standing shock at some fixed radius $r=r_{\rm s}$.
Relative to an unshocked wind with density  $\rho_{\rm w} = \Mdot/4 \pi \vinf r^2$,  the post-shock flow at  $r > r_{\rm s}$ has a density that is a factor $\rho/\rho_{\rm w} = 4$ higher, with a post-shock speed that is a factor $v/\vinf = 1/4$ lower.
The reduction in flow kinetic energy results in a high, immediate post-shock temperature,
\beq
T_{\rm s} = \frac{3}{16} \frac{\mu \vinf^2}{k} = 14 \, {\rm MK} \left ( \frac{\vinf}{1000 \, {\rm km \, s^{-1}}  } \right )^2
\, ,
\label{tsdef}
\eeq
where $k$ is Boltzmann's constant, and  the latter evaluation assumes a standard molecular weight $\mu = 0.62 m_p$,  with $m_p$ the proton mass.

But following this sudden shock increase, the combined effects of adiabatic expansion and radiative cooling cause
the flow temperature $T$ to decrease outward.
For pressure $P= \rho kT/\mu$  and  internal energy density $e=(3/2) P$,
the steady-state energy balance for a general vector velocity ${\bf v}$ is
\beq
\nabla \cdot (e {\bf v}) = - P \nabla \cdot {\bf v} - \rho^2 \Lambda_{\rm m} (T)
\, ,
\label{eneqvec} 
\eeq
where $\Lambda_{\rm m} \equiv \Lambda/(\mu_{\rm e} \mu_{\rm p} )$, with 
$\Lambda(T)$ the optically thin
cooling function \citep[e.g.,][]{Smith01}, and
$\mu_{\rm e} $ and $\mu_{\rm p} $
respectively the mean mass per electron and per proton.

In general, we should also include a detailed momentum equation to account for possible acceleration of the post-shock flow, for example from the inward pull of  stellar gravity, or the outward push of the gas pressure gradient.
But the analysis here is greatly simplified if we make the reasonable assumption that these two countervailing accelerations roughly cancel, and so give a {\em constant speed} $v= \vinf/4$ for all $r > r_{\rm s}$.
For this case of a steady, spherical, constant-speed post-shock outflow, the vector energy equation (\ref{eneqvec}) reduces to a simple differential equation for the decline in temperature with radius $r$,
\beq
\frac{dT}{dr} = - \frac{4}{3} \, \frac{T}{r} - \frac{2\mu}{3kv} \rho  \Lambda_{\rm m} (T)
\, ,
\label{dtdr}
\eeq
wherein the first and second terms on the right-hand-side respectively represent the effects of adiabatic expansion and radiative cooling.
This can alternatively be cast in terms of a temperature scale length,
\beq
\frac{1}{H_{\rm T}} \equiv - \frac{1}{T} \, \frac{dT}{dr} = \frac{4}{3 r} + \frac{\kappa_{\rm c} \rho}{3}
\, ,
\label{htdef}
\eeq
where
\beq
\kappa_{\rm c} (T) \equiv \frac{8 \mu \Lambda_{\rm m} (T)}{ k T \vinf}
\, 
\label{kcdef}
\eeq
is a mass cooling coefficient (with CGS units cm$^2$g$^{-1}$), defined as the {\em inverse} of a characteristic cooling column mass.
The corresponding cooling length  is given by $\ell = 4/\kappa_{\rm c} \rho = 1/\kappa_{\rm c} \rho_{\rm w}$,
defined such that the radiative and adiabatic cooling terms are equal\footnote{The ratio $\ell/r$ differs only by an order-unity factor from the ratio of cooling to escape time, $\chi \equiv  t_{\rm cool}/t_{\rm esc}$, defined by \citet{Stevens92} to characterize the transition from radiative to adiabatic shocks in colliding stellar winds.}
when $\ell = r$. 

\subsection{X-ray luminosity}

For any local post-shock temperature $T$, let $f_{\rm x}(T) $ represent the fraction of  radiation emitted in an X-ray bandpass of interest.
Neglecting for now any wind absorption, the total X-ray luminosity from this single standing shock  is then given by radial integration of the associated X-ray emission,
\beq
L_{\rm xs} =
 4 \pi \int_{r_{\rm s}}^\infty    \rho^2 \Lambda_{\rm m} (T) \, f_{\rm x} (T) \, r^2 dr
\, .
\label{lxrss}
\eeq
Using (\ref{htdef}), this can be recast as an integral over temperature,
\begin{subequations}
\begin{align}
L_{\rm xs} &= 12 \pi \int_{0}^{{T_{\rm s}}}  r^3  \frac{ \rho^2 \Lambda_{\rm m} (T) }{4 + \kappa_{\rm c} (T) \rho r} \, f_{\rm x} (T)  \, \frac{dT}{T} 
\label{lxdtint}
\\
&\approx 48 \pi r_{\rm s}^3  \frac{ \rho_{\rm ws}^2 \Lambda_{\rm m} (T_{\rm s}) }{1 + \kappa_{\rm cs} \rho_{\rm ws}  r_{\rm s}} \, f_{\rm x} (T_{\rm s})  \, \frac{\delta T_{\rm s}}{2T_{\rm s}} 
\, ,
\label{lxrts}
\end{align}
\end{subequations}
where  the latter approximation\footnote{Aside from $f_{\rm x}(T)$, the combination of other factors in the integrand for (\ref{lxdtint}) becomes constant in $T$ in the radiative limit $\kappa_{\rm c} \rho r \gg 1$, and scales as $T^{-3/4}$ in the opposite, adiabatic limit. For $f_{\rm x}(T)$ that declines roughly linearly with temperature, the approximate trapezoidal integration (\ref{lxrts}) thus becomes nearly exact in the radiative limit, while mildly underestimating the actual value in the adiabatic limit, for example by about 15\% for $\delta T_{\rm s} = T_{\rm s}/2$.} uses single-point trapezoidal integration, assuming that $f_{\rm x}$ declines from its post-shock value $f_{\rm x}(T_{\rm s})$ to zero over a temperature range $\delta T_{\rm s}$ from the initial post-shock temperature $T_{\rm s} = T(r_{\rm s})$.
Here  $\rho_{\rm ws} = \Mdot/(4 \pi \vinf r_{\rm s}^2)$ is the wind density just {\em before} the shock, and
\begin{subequations}
\begin{align}
\kappa_{\rm cs} 
\equiv \kappa_{\rm c} (T_{\rm s})
&= \frac{128 \Lambda_{\rm m} (T_{\rm s})}{  3 \vinf^3}
\label{kcsdef}
\\
&= 2 \sqrt{3}  \Lambda_{\rm m} (T_{\rm s})  \left (  \frac{kT_{\rm s}}{\mu } \right )^{-3/2} 
\\
&\approx 1000 \, {\rm \frac{cm^2}{g}} \, T_7^{-2} 
 \approx   750 \, {\rm \frac{cm^2}{g}} \, T_{\rm kev}^{-2} 
\, ,
\label{kcseval}
\end{align}
\end{subequations} 
with $T_7 \equiv T_{\rm s}/10^7 K$ and $T_{\rm kev} \equiv kT_{\rm s}/{\rm keV}$.
The numerical evaluation in (\ref{kcseval}) assumes an approximate fit to the cooling function, $\Lambda (T_{\rm s}) \approx  4.4 \times 10^{-23}/\sqrt{T_7} ~ {\rm erg \, cm^3 s^{-1}}$, over the relevant range of shock temperatures, $10^{6.5} \, K < T_{\rm s} <  10^{7.5} \, K$
\citep{Schure09}.
Recall that the shock cooling length is set by $\ell_{\rm s} = 1/\kappa_{\rm cs} \rho_{\rm ws}$.

For context,
the mass absorption coefficient for bound-free absorption of X-rays
 when smoothed over bound-free edges, also roughly follows an inverse-square scaling with energy.
Over the relevant energy range $0.5-2$~keV, we can use the
 the opacity curves of, e.g., \citet{Cohen10}, \citet{ Leutenegger10}, or  \citet{Herve12} to write an approximate scaling form,
\beq
\kappa_{\rm bf} \approx 30 \, {\rm \frac{cm^2}{g}} \, E_{\rm kev}^{-2} 
\, ,
\label{kbfeval}
\eeq
where $E_{\rm kev}$ is now the X-ray {\em photon energy} in keV.

By casting the cooling strength in an opacity form normally used to describe absorption, we are thus able to make direct comparisons between cooling and absorption, and so characterize their  respective domains of importance. This is further facilitated by the quite fortunate coincidence that both have similar inverse-square scalings with their associated energy.

Since these respective energies are usually roughly comparable, $E_{\rm kev} \approx T_{\rm kev}$,  the fact that the numerical factor for $\kappa_{\rm cs}$ is about 25 times greater than for $\kappa_{\rm bf}$ means that cooling can become important even in winds that are too low density to have significant absorption.  
As such, winds with adiabatic shocks are always optically thin, whereas shocks in optically thick winds are always radiative.
Moreover, as detailed in \S \ref{sec:ewsx},
even in the bulk of O-star winds for which X-ray absorption is weak or marginal, the structure of X-ray emission associated wind shocks should be dominated by radiative cooling.

\subsection{Bridging law between radiative and adiabatic limits}

Noting  that the pre-shock kinetic energy luminosity of the wind is
\beq
L_w = 2  \pi r_{\rm s}^2 \rho_{\rm ws} \vinf^3 = \Mdot \vinf^2/2 \, ,
\label{lwdef}
\eeq
we can use (\ref{kcsdef}) to eliminate the cooling function $\Lambda_{\rm m}$ from (\ref{lxrts}),
recasting the X-ray luminosity scaling as a ``bridging law'' between the radiative and adiabatic shock limits,
\beq
L_{\rm xs} 
=  f_{\rm xs}  \, \frac{9}{16} ~ \frac{ L_w }{1+ \ell_{\rm s}/ r_{\rm s}} 
\, ,
\label{lxblaw}
\eeq
where $f_{\rm xs} \equiv f_{\rm x}(T_{\rm s}) \delta T_{\rm s}/2T_{\rm s}$ is  now a {\em cooling-integrated} shock X-ray fraction.

Note here that the combination of factors multiplying $L_w$ on the right-hand-side of (\ref{lxblaw}) is just the kinetic energy conversion factor introduced in the summary figure \ref{lxlbol-tsm},
\beq
k_{\rm x} \equiv  \frac{9}{16} ~ \frac{ f_{\rm xs}  }{1+ \ell_{\rm s}/ r_{\rm s}} 
\, .
\label{kxdef}
\eeq

For high-density, radiative shocks with $\ell_{\rm s} \ll r_{\rm s}$, 
\beq
L_{\rm xs,rad} = f_{\rm xs}  \, \frac{9}{16} ~ L_w 
= f_{\rm xs}  \, \frac{9}{32} ~ \Mdot \vinf^2
\, .
\eeq
As a physical interpretation, 9/16 is just the fraction of wind kinetic energy that is converted to post-shock heat, which is radiated away before any losses to adiabatic expansion, with the fraction $f_{\rm xs}$ emitted in the X-ray bandpass of interest.
Note that this scales {\em linearly} with density and thus mass loss rate,  
showing that a standard density-squared emission measure does {\it not} represent an appropriate scaling for emission from radiative shocks.

For lower-density, adiabatic shocks with $\ell_{\rm s} \gg r_{\rm s}$, this X-ray emission is reduced by the ratio $r_{\rm s}/\ell_{\rm s}$, giving the scalings,
\begin{subequations}
\begin{align}
L_{xs,ad}  &= f_{\rm xs}  \, \frac{9}{16} \, \frac{r_{\rm s}}{\ell_{\rm s}} \, L_w 
\\
&=  f_{\rm xs} \,  \frac{9  \pi}{8} r_{\rm s}^3  \vinf^3 \kappa_{\rm cs} \rho_{\rm ws}^2 
\\
&= f_{\rm xs}  \, 48  \pi  r_{\rm s}^3  \Lambda_{\rm m} (T_{\rm s}) \rho_{\rm ws}^2 
\label{lxlam}
\\
&= f_{\rm xs} \,  \frac{3  \Lambda_{\rm m} (T_{\rm s}) }{\pi r_{\rm s}} \,  \left (  \frac{\Mdot}{\vinf} \right )^2 
\, ,
\end{align}
\end{subequations}
which thus recovers the density-squared scaling for X-ray luminosity, 
showing that  emission measure does provide an appropriate scaling for emission from adiabatic shocks.

Finally, note that the general bridging law (\ref{lxblaw}) can alternatively be written as a modification of either the radiative or adiabatic scaling,
\begin{subequations}
\begin{align}
L_{\rm xs} &= \frac{L_{\rm xm,rad}}{1+ \ell_{\rm s}/r_{\rm s}} 
\label{lxalt1}
\\
&= \frac{L_{\rm xm,ad}}{1+ r_{\rm s}/\ell_{\rm s}}
\, .
\label{lxalt2}
\end{align}
\end{subequations}
This shows that the X-ray luminosity is always limited to be somewhat {\em below} the {\em smaller} of the radiative or adiabatic luminosities, i.e.\ 
$L_{\rm xs}  \ltwig \min(L_{\rm xm,rad},L_{\rm xm,ad})$.

\subsection{Local X-ray emissivity}
\label{sec:etaxss}

To facilitate application of these single standing-shock scalings to the more complex case of multiple embedded wind shocks generated from the line-deshadowing instability, let us next recast these results in terms of the local X-ray emission from an individual shock.
Since X-ray emission arises from collision of ions and electrons,  it is common to write the  X-ray emissivity (per unit volume and solid angle) as scaling with the square of the local density,
\beq
\eta_{\rm x} = C_{\rm s}  f_{\rm v}  \rho^{2}
\, ,
\label{etaxdef}
\eeq
where
$C_{\rm s}$ is a constant that depends on the shock strength and atomic physics,
and $f_{\rm v}$ represents a local volume filling factor for shocked gas that is sufficiently hot to emit X-rays.
If each {\em individual} shock has an associated filling factor $f_{\rm vs}$ 
proportional to its post-shock cooling length, 
then the total filling factor from an {\em ensemble} of shocks can be written,
\beq
f_{\rm v} (r)  = f_{\rm vs} \, \frac{dN_s}{d\ln r} = f_{\rm vs} \, n_{\rm s} (r) 
\label{fvdndef}
\, ,
\eeq
where $N_s (r)$ is the {\em cumulative} number of shocks up to radius $r$,  and $n_{\rm s}$ measures the local {\em differential} number of new, emerging shocks.
This formalism emphasizes the importance of the number of shocks and their spatial distribution;
compared to the traditional emission measure approach,
it should provide more physically motivated constraints for wind-shock X-ray production in massive stars.

In particular, the analysis below of multiple, instability-generated shocks assumes a power-law scaling for $n_{\rm s}$ (see eq.\ \ref{fqpowlaw}); but for the above single-shock model, this just takes the form of a Dirac delta-function, $n_{\rm s} = r \delta (r-r_{\rm s})$.
The associated X-ray luminosity is then given by integration of the emissivity $\eta_{\rm x}$ over solid angle and volume,
\beq
L_{\rm x} = 16 \pi^2 \int  C_{\rm s} f_{\rm vs}  \, r \delta(r-r_{\rm s}) \rho^2 \, r^2 dr
\, .
\label{lxthindef}
\eeq
Upon trivial evaluation over the delta function, comparison with the scalings (\ref{lxlam}) and (\ref{lxalt2}) yields the identifications,
\beq
C_{\rm s} = \frac{3}{\pi} \Lambda_{\rm m} (T_{\rm s}) f_{\rm xs}
\,
\label{csdef}
\eeq
and 
\beq
f_{\rm vs} = \frac{1}{1 + r_{\rm s}/\ell_{\rm s}}
\, .
\label{fvsdef}
\eeq
This thus now sets a bridging law at the level of an individual shock, with $f_{\rm vs}$ characterizing the fraction of single-shock emission measure that actually contributes to radiative emission.
The adiabatic limit $\ell_{\rm s} \gg r_{\rm s}$ gives $f_{\rm vs} \approx 1$ and so a density-squared scaling for the emissivity (\ref{etaxdef}); the radiative limit reduces this by the small factor $\ell_{\rm s}/r_{\rm s} \ll 1$, thus giving this emissivity a linear-density scaling set by the total kinetic energy flux through the shock.

Note that the factor $C_{\rm s}$ depends only on the shock strength (through the post-shock temperature $T_{\rm s}$), while the shock filling factor $f_{\rm vs}$ depends on the local shock radius and cooling length, as set by the pre-shock wind density.
Thus, although these scalings are derived from a simple model of a single, standing shock, they should also be generally applicable to more complex, moving shock structures, if specified in terms of the pre-shock wind density\footnote{For instability-generated wind structure, X-rays can arise from collisions between clumps that have been compressed to some fraction of the wind volume \citep{Feldmeier97}, implying then a higher input density that would lower the cooling length. For simplicity, the analysis here does not account for this possibility, but it could enhance the importance of radiative cooling, leading to an even larger effective ratio $\kappa_{\rm cs}/\kappa_{\rm bf}$ of cooling to absorption.} and the relative velocity jump that sets the shock strength and temperature.

\section{X-rays from Instability-Generated Embedded Wind Shocks}
\label{sec:ewsx}

\subsection{Exospheric scaling for $L_{\rm x}$}

To model the X-ray emission from multiple, embedded wind shocks, let us next write the X-ray luminosity in a fully general, 3-D form that accounts for possible directional dependencies in emission and absorption,
\beq
L_{\rm x} = \int d^3 r \int d\Omega \,  \eta_{\rm x}({\bf r},{\bf n} ) \, e^{-\tau({\bf r},{\bf n} )}
\, ,
\label{lxgen}
\eeq
where the optical depth $\tau({\bf r},{\bf n} )$ accounts for bound-free absorption of X-rays emitted at location ${\bf r}$ in the observer direction ${\bf n}$.

In principle, instability-generated, X-ray emitting  shocks will be associated with a complex, 3D, stochastic wind structure
\citep{Dessart03}.
But upon averaging over small scales,  this can be reasonably well described by a globally spherical wind emission model (OC99).
In modeling line emission, accounting for the observed Doppler shift from wind expansion still requires including a directional dependence of the emissivity and optical depth; see \S \ref{xlinesec} and  \citet[][hereafter OC01]{Owocki01}.
But to derive the general scalings for the  X-ray luminosity, one can again take the total X-ray emissivity to be an {\em isotropic} function of the local radius, $\eta_{\rm x} (r)$.
Moreover, given the weak to moderate importance of absorption in all but the densest winds, its overall role in scaling relations can be roughly taken into account through a simple {\em exospheric} approximation \citep[OC99,][]{Leutenegger10}, for which the integrations over solid angle and volume in (\ref{lxgen}) reduce to just a single integration in radius,
\beq
L_{\rm x} \approx  16 \pi^2 \int_{R_{\rm i}}^\infty \, 
\eta_{\rm x} (r) \,   r^2 dr
\label{lxetaexo}
\eeq
where the integral lower bound is taken from the {\em larger} of the X-ray onset radius and the radius for unit radial optical depth, i.e.\  $R_{\rm i} \equiv \max [R_{\rm o},R_{\rm 1}]$.
For a wind with mass loss rate $\Mdot$ and flow speed $\vinf$, this radius for  transition from optically thick to thin X-ray emission is given by
\beq
R_{\rm 1}  
\equiv \tau_{\ast} \Rstar 
= \frac{\kappa_{\rm bf} \Mdot}{4 \pi \vinf }
\approx 25  R_{\odot} \, \frac{\Mdot_{\rm -6}}{E_{\rm kev}^{2} V_{\rm 1000} }
\, ,
\label{r1def}
\eeq
where  $\tau_\ast$ is a characteristic wind optical depth  to the stellar surface radius $\Rstar$,
and the latter equalities use the scaling (\ref{kbfeval}) for the bound-free X-ray opacity $\kappa_{\rm bf}$, 
with $\Mdot_{\rm -6} \equiv \Mdot/(10^{-6} M_{\odot}$~yr$^{-1}$)
and
$V_{\rm 1000} \equiv \vinf/1000 \,$km~s$^{-1}$.

In direct analogy,
we can similarly define a characteristic {\em adiabatic radius} for transition from radiative to adiabatic cooling of the associated wind shocks,
\beq
R_{\rm a}  \equiv  \frac{\kappa_{\rm cs} \Mdot}{4 \pi \vinf } 
\approx
625  R_{\odot} \, \frac{\Mdot_{\rm -6}}{T_{\rm kev}^{2} V_{\rm 1000} }
\, ,
\label{radef}
\eeq
where $T_{\rm kev} = kT_{\rm s}/$keV.
For X-rays emitted with  photon energy comparable to the shock energy, $E_{\rm kev} \approx T_{\rm kev}$, we find $R_{\rm a}/R_1 \approx \kappa_{\rm cs}/\kappa_{\rm bf} \approx 25$. As detailed below, this implies there can be extensive wind regions in which the density is high enough for shocks to be radiative, but too low for much wind absorption of their emitted X-rays. 
In particular, independent of the mass loss rate, any shocks formed near the radius $R_1$, where X-rays have near unit  optical depth, should always be strongly radiative.

\begin{figure}
\includegraphics[angle=0,scale=0.4]{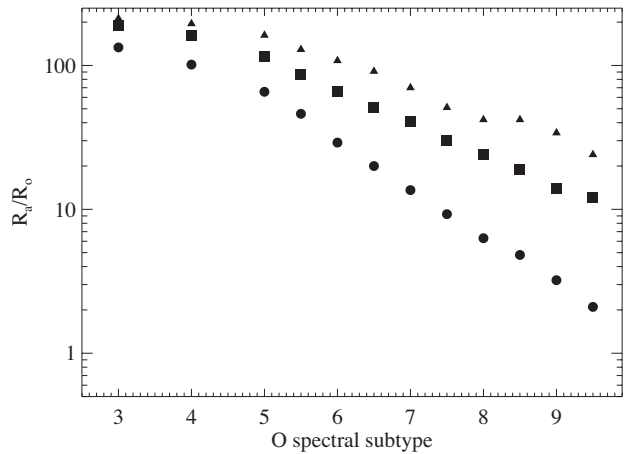}
\caption{Ratio $R_{\rm a}/R_{\rm o}$ of adiabatic radius to shock onset radius, assuming $R_{\rm o} = 1.5 \Rstar$, and a post-shock temperature  $T_{\rm kev} = 0.5$.
To construct the plot, stellar parameters were taken from Tables 1-3 of \citet{Martins05} ($T_{\rm eff}$, $L_{\rm bol}$, $\Rstar$, $\Mstar$) for luminosity classes V, III, and I (marked respective by circles, squares, and triangles). The terminal speeds are computed as $\vinf =  2.6 V_{\rm esc} $, and mass loss rates are from \citet{Vink00}.
}
\label{raro}
\end{figure}

Figure \ref{raro} plots estimates of $R_{\rm a}/R_{\rm o}$ vs. O-star spectral subtype, for shocks with the  temperature  $T_{\rm kev} = 0.5$, as expected for wind-shocks generated by the line-deshadowing instability \citep{Runacres02, Dessart03}, and estimated from the observed, relatively soft X-ray spectra \citep{Wojdowski05, Zhekov07, Naze12}. 
The large $R_{\rm a}$ values confirm that  wind-shocks of this energy should be radiative.

This plot can also be readily used to estimate expected unit optical depth radii.  In particular, for an emission line at $E_{\rm kev}=1$, the corresponding $R_1$ would be a factor 100 smaller than the $R_{\rm a}$ plotted in figure \ref{raro}.  Since this means most O-stars would have $R_1/R_{\rm o} \ltwig 1$, we see that absorption effects should be weak to moderate throughout the O-star domain, even while the shocks are generally radiative. 
See \S \ref{sec:fulllxin} and figure \ref{lxvsrab1} for further discussion and illustration of these scalings.

\subsection{Bridging law 
with thin-shell mixing}

Previous analyses using the density-squared emissivity (\ref{etaxdef})
(e.g., OC99, OC01) 
have  directly parameterized the X-ray filling factor  $f_{\rm v}$ as following some specified radial function, e.g.,  a power-law.
But instead let us now parameterize the shock {\em number} distribution in (\ref{fvdndef}) by an analogous power-law,
\beq
n_{\rm s} (r) \equiv  n_{\rm so} \left ( \frac{R_{\rm o}}{r} \right )^{\q} ~~ ; ~~ r > R_{\rm o}
\, ,
\label{fqpowlaw}
\eeq
with $n_{\rm s}=0$ for $r < R_{\rm o}$. 
Both instability simulations \citep{Owocki99, Runacres02,  Dessart03} and 
X-ray profile fitting and He-like f/i ratios \citep{Leutenegger06, Cohen06, Cohen10, Cohen11}
suggest an initial onset for shock formation around $R_{\rm o} \approx 1.5 \Rstar$.

Applying (\ref{fqpowlaw}) in (\ref{fvdndef}), the analysis in \S\ref{sec:etaxss}  provides a more physical model of shock X-ray emission that accounts explicitly for the effects of radiative cooling.
However, it still does not account for any thin-shell {\em mixing}.
The inherent thinness ($\ell \ll r$) of radiative shock cooling zones makes them
subject to various thin-shell instabilities 
\citep{Vishniac94}, 
which in numerical simulations lead to highly complex, turbulent shock structure 
\citep[e.g.,][]{Walder98}.
\citet{Parkin10} discuss how the inherently limited spatial resolution of radiative shock simulations leads to a ``numerical conduction'' that transports heat from high to low temperature gas, resulting in a severe, but difficult-to-quantify reduction in the X-ray emission.

While the specific mechanisms within hydrodynamical simulations may indeed depend on such {\em numerical} artifacts, the perspective advocated here is that such an overall reduction in X-ray emission is likely a natural consequence of the turbulent cascade induced by the thin-shell instability; 
this should lead to  substantial {\em physical} mixing between cool and hot material, with the softer and more efficient radiation of the cooler gas effectively reducing the emission in the X-ray bandpass.

Pending further simulation studies to quantify such mixing and X-ray reduction, we  make here
the plausible {\em ansatz} that the reduction should, for shocks in  the radiative limit $\ell/r \ll 1$, scale as some power $m$ of the cooling length ratio, $(\ell/r)^m$.
To ensure that the 
mixing becomes ineffective in the adiabatic limit -- for which the cooling layer is too extended to be subject to thin-shell instability -- 
we recast the bridging law (\ref{fvsdef}) for the shock volume filling factor $f_{\rm vs}$ in the generalized form,
\beq
 f_{\rm vs}  \approx  \frac{1}{ ( 1+r/\ell )^{1+m}}
 =  \frac{1}{ ( 1+ \kappa_{\rm c} \rho r )^{1+m}} 
  \, ,
\label{fvmbridge}
\eeq
with the level of mixing now controlled by a positive value of the mixing exponent $m$. 
To simplify notation in the analysis to follow, we have dropped here the subscripts (``s" or ``w") for the quantities (e.g., $\ell$, $r$, $\kappa_{\rm c}$, $\rho$) on the right-hand-side.
Recalling that the cooling coefficient $\kappa_{\rm c}$ takes the scalings given in  equations (\ref{kcsdef})-(\ref{kcseval}), we can 
use the adiabatic radius $R_{\rm a}$ from (\ref{radef}) to characterize the asymptotic regimes for (\ref{fvmbridge}).

If $R_{\rm a} < R_{\rm o}$, then the shocks are adiabatic throughout the wind,  and we recover the standard density-squared emissivity (\ref{etaxdef}) with a specified volume filling factor set by $f_{\rm v} = n_{\rm s}$.

If $R_{\rm a} > R_{\rm o}$, this adiabatic scaling still applies in the outer wind, $r > R_{\rm a}$;
but in the inner regions  $R_{\rm o} < r < R_{\rm a}$, the cooling term in the denominator of (\ref{fvmbridge}) dominates, giving the emissivity (\ref{etaxdef}) now a {\em reduced} dependence on density,
\beq
\eta_{\rm x} 
\approx C_{\rm s} n_{\rm s}  \frac{\rho^{1-m}}{(\kappa_{\rm c} r)^{1+m}} ~~ ; ~~ R_{\rm o} < r < R_{\rm a} 
\, .
\label{etaxrad}
\eeq
Without mixing ($m=0$), the density dependence is thus linear, but with mixing ($m>0$), it becomes {\em sub-linear}.

As noted above,  absorption is a modest effect in even dense O-star winds, with $\tau_\ast$ at most of order unity, implying then that $ R_1 \ltwig R_{\rm o}$
\citep{Cohen10}.
But the stronger coefficient ($\kappa_{\rm c}/\kappa_{\rm bf} = R_{\rm a}/R_1 \approx 25$) for radiative cooling means that such winds typically have $R_{\rm a}  > R_{\rm o}$, implying that most  shocks remain radiative well above the wind acceleration region where they are generated.

\subsection{$L_{\rm x}$ scalings with $\Mdot/\vinf$}
\label{sec:lxscale}

Let us now turn to the scalings for the overall X-ray luminosity. 
Applying the emissivity (\ref{etaxdef}) and filling factor (\ref{fvmbridge}) to the exospheric model (\ref{lxetaexo}) with the power-law shock distribution (\ref{fqpowlaw}), we find
\begin{align}
L_{\rm x}  &\approx 16 \pi^2  C_{\rm s} \int_{R_{\rm i}}^\infty \, 
 \frac{n_{\rm s} \rho^{2}}{( 1+\kappa_{\rm c} \rho r )^{1+m}} \,  r^{2} dr
\nonumber
\\
 &= 
 \cq
\left [ \frac{\Mdot}{\vinf} \right ]^{2} 
\int_{R_{\rm i}}^{\infty} \frac{ dr}{r^{\q} (rw)^{1-m} (rw+R_{\rm a})^{1+m}} \,
\, ,
\label{lxri}
\end{align}
where $\cq \equiv C_{\rm s} n_{\rm so} R_{\rm o}^{\q}$,
and the latter equality uses mass conservation to cast 
the integral in terms of
the scaled wind velocity,  $w(r) \equiv v(r)/\vinf$.
This  is generally taken to have a `beta-law' form  $w(r) = (1-\Rstar/r)^\beta$, with 
the canonical case $\beta=1$ giving $rw = r- \Rstar$, and the constant-speed case $\beta=0$ giving $rw=r$.
For these special cases, \S \ref{sec:fulllxin} gives some  results for full integrations of (\ref{lxri}).

But even for a generic velocity law $w(r)$, we can readily glean the essential scalings of the $L_{\rm x}$ from (\ref{lxri}) in key asymptotic limits of adiabatic vs.\ radiative shocks. 

\subsubsection{Adiabatic shocks in optically thin wind}

First for low-density winds with optically thin emission from adiabatic shocks, $R_1 <  R_{\rm a} < R_{\rm o}$, we can effectively drop the $R_{\rm a}$ term in the denominator, and set the lower bound of the integral to the fixed onset radius, $R_{\rm i} = R_{\rm o}$,
yielding
\beq
L_{\rm x} \approx \cq
\left [ \frac{\Mdot}{\vinf} \right ]^{2} 
\int_{R_{\rm o}}^{\infty} \frac{ dr}{w^2 \, r^{\q+2} }  \, ~~ ;  ~~ R_1 < R_{\rm a} < R_{\rm o}
\, .
\label{lxad}
\eeq
Since the resulting integral then is just a fixed constant that is independent of $\Mdot$, the X-ray luminosity recovers the standard adiabatic scaling $L_{\rm x} \sim (\Mdot/\vinf)^2$.

\subsubsection{Radiative shocks in optically thin or thick wind}

For high-density winds with radiative shocks and so $ R_{\rm o} < R_{\rm a}$,
the $R_{\rm a}$ now dominates its term, and so can be pulled outside the integral.
Rescaling the remaining integrand in terms of the initial radius $R_{\rm i}$, we find
\beq
L_{\rm x} = \cq
\left [ \frac{\Mdot}{\vinf} \right ]^{2} \frac{R_{\rm i}^{m-\q}}{R_{\rm a}^{m+1}}
\int_{1}^{\infty} \frac{ dr}{r^{\q} (rw)^{1-m}}  \, ~~ ;  ~~  R_{\rm o} < R_{\rm a}
\, ,
\label{lxrad}
\eeq
where again the integral is now essentially independent of $\Mdot/\vinf$.

For the intermediate-density case in which the radiative shock emission is optically thin, $R_1 < R_{\rm o} < R_{\rm a}$, the integration lower limit is fixed at the onset radius, $R_{\rm i} = R_{\rm o}$, which is independent of $\Mdot/\vinf$. But since $R_{\rm a} \sim \Mdot/\vinf$, the overall scaling is  $L_{\rm x} \sim (\Mdot/\vinf)^{1-m}$.
 
For the case of very dense, optically thick winds with radiative shocks, $R_{\rm o} < R_{\rm 1} < R_{\rm a}$, the lower boundary at $R_{\rm i} = R_{\rm 1}$ gives the residual integral an additional dependence on $R_{\rm 1}^{m-\q}$; since $R_{\rm 1}$ too scales with mass loss rate, the dependence on the mixing index $m$ {\em cancels}. 
At the radius $R_{\rm 1}$ the cooling length ratio $\ell/r$ is always the same, implying that in optically thick winds the observed radiative shock emission is likewise fixed for any mass loss rate.
The luminosity scaling thus becomes {\em independent} of mixing, scaling just with shock number index as $L_{\rm x} \sim (\Mdot/\vinf)^{1-\q}$.

This is in fact the same scaling found in OC99 for optically thick winds, but with {\em adiabatic} shocks. Indeed, OC99 argued that  
assuming a filling factor $f_{\rm v} \sim r^{-0.4}$
could give a sub-linear dependence on mass loss rate, $L_{\rm x} \sim \Mdot^{0.6}$, and so possibly 
reproduce
 the $L_{\rm x} \sim L_{\rm bol}$ relation.
Subsequent analysis of X-ray line profiles observed from {\it Chandra} and {\it XMM-Newton} have shown, however, that optical depth effects are quite moderate even for O-stars like $\zeta$~Puppis with quite dense winds
\citep{ Cohen10}.
The bulk of O-star winds are simply too low-density for this optical thickness scaling to apply, and so this cannot be the explanation for the $L_{\rm x}  \sim  L_{\rm bol}$ relation.

\subsubsection{Summary of asymptotic scalings}

To summarize, power-law shock-number models with thin-shell mixing have the asymptotic scalings,
\begin{subequations}
\begin{align}
L_{\rm x} 
&\sim \left [ \frac{\Mdot}{\vinf} \right ]^{2} 
~~~ ; ~ R_{\rm 1} < R_{\rm a} <  R_{\rm o}  ~;~ {\rm adiabatic,  ~ thin}
\label{lxmd1}
\\
&\sim  \left [
\frac{\Mdot}{\vinf }  
\right ] ^{1-m}  ; ~ R_{\rm 1} < R_{\rm o} < R_{\rm a}   ~;~ {\rm radiative, ~ thin}
\label{lxmd2}
\\
&\sim \left [ \frac{\Mdot}{\vinf} \right ]^{1-\q} \,  ~;~ R_{\rm o} < R_{\rm 1} < R_{\rm a}   ~;~ {\rm radiative, ~ thick}
\label{lxmd3}
\end{align}
\end{subequations}
where the progression represents a trend of increasing  $\Mdot/\vinf$.
The first applies for weak winds, for example from early B main-sequence stars, which typically show weak X-ray emission, with $L_{\rm x}$ well below $10^{-7} L_{\rm bol}$
\citep{Cohen97}.
The middle scaling for intermediate-density winds is the most relevant for the bulk of O-type stars found to follow the $L_{\rm x} \sim L_{\rm bol}$ relation.
The last applies only to the strongest winds, e.g.\ very early O supergiants like HD93129A, for which analysis of X-ray line profiles show moderate absorption 
effects\footnote{Indeed, this star could be viewed as a transitional object to the WNH type Wolf-Rayet stars,  for which absorption effects should strongly attenuate any X-rays from instabilities in the wind acceleration region; see \S \ref{sec:summary}.}, with $\tau_\ast \gtwig 1$ \citep{ Cohen11}.

\subsection{Link between $\Mdot$ and $L_{\rm bol}$ scaling}
\label{md2lbol}

As noted in the introduction, and summarized in figure \ref{lxlbol-tsm}, straightforward application of CAK wind theory implies a {\em direct} dependence of mass loss rate on luminosity that scales as $\Mdot \sim L_{\rm bol}^{1/\alpha}$, where $\alpha \approx 0.6$ is the CAK power index; for the bulk of O-type stars with intermediate-density winds and thus $L_{\rm x} \sim \Mdot^{1-m}$, reproducing the observed $L_{\rm x} \sim L_{\rm bol}$ relation thus simply requires a mixing exponent $m \approx 1- \alpha \approx 0.4$.

More generally, let us now  consider how this requirement is affected if one accounts also for a secondary dependence on stellar mass $M$,
 which in turn can give a further {\em indirect} dependence on $L_{\rm bol}$.
 Specifically, within CAK  wind theory, $\Mdot \sim M^{1-1/\alpha}$ and $\vinf \sim M^{1/2}$, and so if we in turn assume from stellar structure a mass-luminosity dependence $M \sim L_{\rm bol}^s$, where $s \approx 1/3$ \citep[e.g.,][p. 360]{Maeder09}, we find

\beq
\log \left ( \frac{\Mdot}{\vinf} \right ) 
\sim \frac{2 - s  (2- \alpha) }{2\alpha } \, \log(L_{\rm bol})
\sim \frac{ \log (L_{\rm x} )}{1-m}
\, ,
\eeq
where the latter relation makes use of  the scaling (\ref{lxmd2})\footnote{Of course, 
the terminal speed also depends on stellar radius as $\vinf \sim 1/\sqrt{\Rstar }$, but  the diverse luminosity classes of X-ray emitting O-stars makes it difficult to identify any {\em systematic} dependence on $L_{\rm bol}$ that might influence the overall $L_{\rm x}  - L_{\rm bol}$ relation.}.

Reproducing the  empirical $L_{\rm x} \sim L_{\rm bol}$ relation thus now requires
\beq
m =  \frac{2 (1 -\alpha)  - s (2 - \alpha )} {2 - s (2 - \alpha )} 
, .
\eeq
Accounting for a stellar structure scaling $s \approx 1/3$ with a CAK index $\alpha \approx 0.6$ thus now requires a mixing exponent $m \approx 0.22$, somewhat smaller than the $m \approx 0.4$ required if one assumes no systematic mass-luminosity scaling (i.e., $s=0$).

\begin{figure}
\includegraphics[angle=0,scale=0.33]{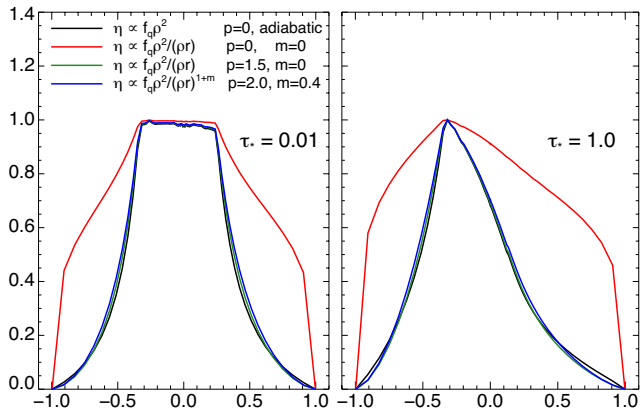}
\caption{
X-ray line profiles (normalized to unit maximum) vs. Doppler-shifted wavelength (normalized to shift for terminal velocity $\vinf$), for optically thin ($\tau_\ast \ll 1$; left) and marginally optically thick ($\tau_\ast =1$; right) lines.
The black curves show the adiabatic, constant-filling-factor models that give good general fits to observed X-ray emission lines. The other curves show radiative models for no mixing and for $m=0.4$, with shock-number indices $\q$ as labeled. Radiative models with constant shock number ($\q=0$) give profiles that are too broad (red curves). But models with steeper number exponents ($\q=1.5$ for $m=0$ in green; and $\q=2$ for $m=0.4$ in blue) fit well the observationally favored black curves.
If the O-star $L_{\rm x} - L_{\rm bol}$ relation is to be 
reproduced
with moderate thin-shell mixing ($m \approx 0.4$) of radiative shocks, then matching observed X-ray emission lines requires the shock number to have a moderately steep decline ($\q \approx 2$) above the initial onset radius $R_{\rm o}$.
}
\label{lxprofiles}
\end{figure}

\section{Effect of Shock Cooling on X-ray Line Profiles}
\label{xlinesec}

\subsection{Basic formalism}

Observations by {\it XMM-Newton} and {\it Chandra} of spectrally resolved, wind-broadened X-ray emission lines from luminous OB stars provide a key diagnostic of the spatial distribution of X-ray emission and absorption in their expanding stellar winds
\citep{Ignace02, Oskinova06, Owocki06}.
In particular, the relatively weak blue-red asymmetry of the observed emission lines 
indicates
modest wind optical depths, $\tau_\ast \sim 1$, while the overall width constrains the spatial location of the emission within the expanding velocity law.
For the usual density-squared emission model with a prescribed (power-law) spatial variation in volume filling factor $f_{\rm v}$, fits to observed X-ray lines 
are typically consistent with
 a standard $\beta \approx 1$ velocity law and an $f_{\rm v}$ that  is spatially nearly {\em constant}, corresponding to $\q \approx 0$ within the adiabatic scaling implicit in the emission measure analysis
\citep{Kramer03, Cohen06}.
In the discussion below, we refer to these profiles -- plotted in black in figure \ref{lxprofiles} -- as ``observationally favored".

Within the perspective discussed here that shocks within most O-star winds are likely to be radiative instead of adiabatic, let us now examine how inclusion of  radiative cooling affects X-ray line profiles.
Following \citet[hereafter OC01]{Owocki01}, 
the directional Doppler-shift of the X-ray line emission within the expanding wind is modeled through a
line emissivity $\eta_\lambda (r,\mu)$ at an observer's wavelength $\lambda$ along
direction cosine $\mu$ from a radius $r$.
The resulting X-ray luminosity spectrum $L_\lambda$ is computed from
integrals of the emission over direction and radius, attenuated by
bound-free absorption within the wind (cf. OC01 equation 1),
\begin{equation}
L_\lambda = 8 \pi^2 \int_{-1}^1 \,d \mu \, 
\int_{\Rstar}^\infty \, dr \, r^2 \eta_\lambda (\mu,r) \, e^{-\tau (\mu,r)} .
\label{llamdef}
\end{equation}
\noindent
The absorption optical depth $\tau (\mu,r)$ is
evaluated by converting to ray coordinates and then integrating for each ray with a fixed impact 
parameter  from the local position to the observer.
For the standard $\beta = 1$ velocity law, 
the integrals are analytic,
 with overall  scaling in proportion to $\tau_\ast \equiv R_1/R_\ast$.

 In principle, this integrated optical depth can be affected by the ``porosity'' associated with optically thick clumps or anisotropic ``pancakes''  \citep{Feldmeier03}, with potential consequences for interpreting the {\em asymmetry} of  X-ray line profiles in terms of the wind mass loss rate
 \citep{Oskinova06}.
But for  the optically thin ($\tau_\ast \ll 1$) or marginally optically thick ($\tau_\ast  \sim 1$) lines considered here,  individual clumps should be optically thin \citep{Owocki06, Sundqvist12b}, and so such effects are not important for the discussion below, which focuses on the overall profile width.

%
 
\subsection{Scaling analysis}

As noted, applications of this OC01 formalism assuming a density-squared emission 
show that a spatially constant X-ray volume filling factor $f_{\rm v}$ gives generally quite good fits to observed X-ray lines.
To examine the effect of radiative cooling, let us now apply the more general bridging-law scalings of (\ref{etaxdef}) and (\ref{fvmbridge}), assuming the simple power-law form for the shock number (\ref{fqpowlaw}).
The spatial integration thus takes the same form as the exospheric result (\ref{lxri}), except that  absorption is now treated explicitly by the exponential optical depth term in the integrand, with the radial lower bound fixed at the X-ray onset radius, $R_{\rm i} = R_{\rm o}$.

We can again infer  basic scaling results by inspection of this integrand in the limits of adiabatic vs.\ radiative shocks.
For the adiabatic case, this follows the scaling in (\ref{lxad}), with radial dependence $1/w^2 r^{\q+2}$. To match observed profiles, such adiabatic emission models require constant $f_{\rm v}=n_{\rm s}$, with the zero power-law exponent $\q=0$ implying a $1/(wr)^2$ variation of the integrand.

By contrast,  the radiative limit follows the scaling in (\ref{lxrad}), with direct radius dependence $1/r^{\q+1-m}$, and velocity dependence $1/w^{1-m}$ that is weaker than the $1/w^2$ of the adiabatic model.  Focusing first just on the former, we see that reproducing the $1/r^2$ integrand needed to fit observed profiles would now require $\q=1+m$.  In practice, to compensate for the weaker inverse-speed dependence, fitting the observed profile width requires a somewhat steeper shock-number decline, as we now quantify.

\subsection{X-ray line profiles from radiative shocks} 

For a $\beta=1$ velocity law with $R_{\rm o}=1.5 \Rstar$ and various specified values of the exponents $\q$ and $m$,
figure~\ref{lxprofiles} plots normalized X-ray line profiles for optically thin ($\tau_\ast \ll 1$; left)  and marginally optically thick ($\tau_\ast = 1$; right) lines.
In both cases, the black curve represents the adiabatic, constant-filling-factor ($\q=0$) model that gives generally good fits to observed profiles.

The other curves show results for radiative shocks. 
Without mixing, the red curve for radiative shocks with constant shock number ($\q=0$) is far too broad;
fitting the favored black profile now requires a $\q=1.5$ (blue curve), 
which is even steeper  than the predicted $\q=m+1=1$ needed to compensate for the weaker direct radial scaling.
With mixing exponent of $m=0.4$, we find a $\q=1.5$ shock-number model  
gives profiles 
(not shown) 
that are still somewhat too broad.
But with a somewhat steeper number exponent $\q=2$, the blue curve again nearly reproduces the black curve\footnote{Also not shown here are profiles computed for $\q=2$  and the alternative mixing exponent value $m = 0.22$, which we find also give close agreement with the black curves.}.

An overall conclusion is thus that, for radiative shock models with mixing at a level needed to 
reproduce
the $L_{\rm x} \sim L_{\rm bol}$ relation, matching observed X-ray emission requires a steep radial decline  ($\q \approx 2$) in shock number above the onset radius.

\begin{figure*}
\begin{center}
\includegraphics[scale=1.75]{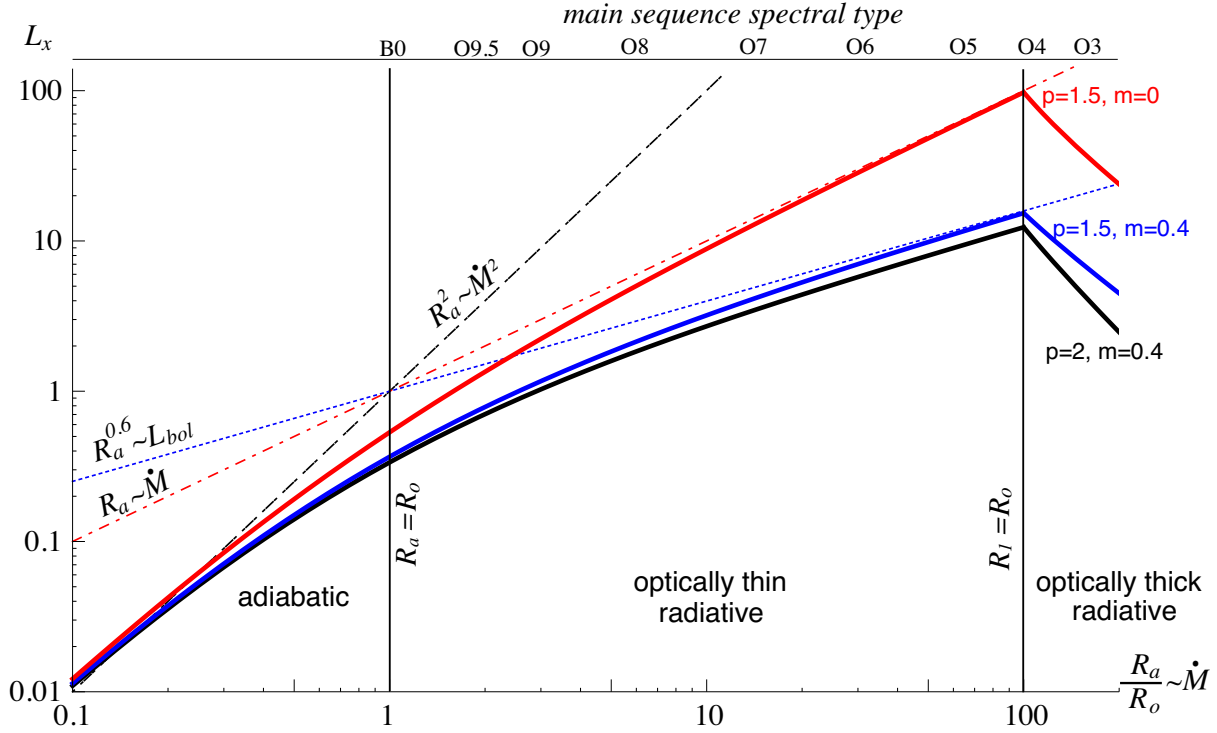}
\caption{
 Normalized X-ray luminosity $L_{\rm x}$ vs. adiabatic radius $R_{\rm a}$ scaled by shock onset radius $R_{\rm o} = 1.5 \Rstar$, for winds with a standard $\beta=1$ velocity law, 
 and selected, labelled values for the parameters $\q$ and $m$ . 
 The dashed, dot-dashed, and dotted lines show the expected scalings for respectively adiabatic, radiative, and thin-shell-mixed shocks.
As discussed in the text, the variation in $R_{\rm a}$ represents a proxy for the wind density parameter $\Mdot/\vinf$, ranging from low density, adiabatic shocks on the far left, to high-density, optically thick winds with radiative shocks on the far right;
the optically thick turnover at $R_{\rm a}/R_{\rm o} > 100$ applies for observed X-ray energies that are about a factor two higher than the shock energy.
The intermediate density case with optically thin, radiative shocks follows the $L_{\rm x} \sim \Mdot^{1-m}$ scaling that reproduces the empirical $L_{\rm x} \sim L_{\rm bol}$ relation if $m \approx 0.4$.
The upper axis uses the analysis from figure \ref{raro} to mark the corresponding spectral class for main sequence stars;
for higher luminosity giants and supergiants, the associated spectral class sequence would shift to the right.
}
\label{lxvsrab1}
\end{center}
\end{figure*}

\subsection{Decline of $L_{\rm x}$ in optically thick winds}

Such a steep radial drop off in shock number has important
implications for the scaling of X-ray luminosity for the highest density stars that become optically thick to bound-free absorption, i.e. with $\tau_\ast > 1$ and thus $R_1 > R_{\rm o}$.
Namely, from the scaling given by (\ref{lxmd3}), we see that taking $\q \approx 2 $ implies that the X-ray luminosity for such optically thick winds should now {\em decline inversely} with mass loss rate, $L_{\rm x} \sim \Mdot^{1-\q} \sim 1/\Mdot$.
If the X-ray emission is concentrated near an onset radius within the wind acceleration zone, the bound-free absorption by the overlying, optically thick wind significantly attenuates the net X-rays seen by an external observer.
For early O-type supergiants with dense winds, this can lead to a reduced X-ray luminosity,  but because the overall decline of bound-free opacity with X-ray energy, the observed spectrum can be hardened.
The recent analysis of X-rays from the O2If star HD93129A provides a potential example approaching this limit
\citep{Cohen11}.

\section{Scaling results for full evaluation of  $L_{\rm x}$ integral}
\label{sec:fulllxin}

\subsection{Constant-speed wind with $\beta=0$}

As a supplement to the asymptotic $L_{\rm x}$ scalings given in  \S \ref{sec:lxscale}, let us finally consider full solutions for the general integral (\ref{lxri}).
For a constant speed model with $\beta=0$ and thus $w(r)=1$, general analytic integration is possible in terms of the incomplete beta function.
But the typical properties can be more simply gleaned by examining the special case $\q=1$, for which the integral in (\ref{lxri})  takes the simpler analytic form,
\beq
L_{\rm x} = \frac{\cq \q \kappa_{\rm c}^2}{16 \pi^2 m (1-m) } \, 
\left [ \frac{1 + m R_{\rm a}/R_{\rm i}}{(1+R_{\rm a}/R_{\rm i})^m} -1 \right ] \, ,
\label{lxb0q0}
\eeq
wherein the square-bracket factor sets the scalings with $\Mdot/\vinf$, with  the preceding terms just fixing the overall normalization.
 For low-density, optically thin winds, the initial radius is fixed to the onset radius, so that
 $R_{\rm a}/R_{\rm i}  \sim R_{\rm a}/R_{\rm o} \sim \Mdot/\vinf $.  For $R_{\rm a}/R_{\rm o} \ll 1$, expansion of the square-bracket term recovers the adiabatic scaling $L_{\rm x} \sim (\Mdot/\vinf)^2$ of (\ref{lxmd1}), while for $R_{\rm a}/R_{\rm o} \gg 1$, it becomes proportional to $R_{\rm a}^{1-m}$ and so recovers the radiative, optically thin scaling (\ref{lxmd2}).
 For high-density, optically thick winds, the ratio $R_{\rm a}/R_{\rm i} = R_{\rm a}/R_1$ becomes independent of mass loss rate, and so the square-bracket factor and thus $L_{\rm x}$ approach a constant value, in accord with (\ref{lxmd3}) for this case with $\q=1$.
 
\subsection{Standard wind with $\beta=1$}

For a standard  $\beta=1$ velocity law with $rw = r - \Rstar$, analytic integration of (\ref{lxri}) can  be cast in terms of the Appell hypergeometric function; but in practice it is more straightforward just to carry out the integration numerically. 

Figure \ref{lxvsrab1} plots the resulting X-ray luminosity $L_{\rm x}$  (scaled by the dimensional factor outside the square brackets in (\ref{lxb0q0})) vs.\ adiabatic radius $R_{\rm a}$ (scaled by the shock-onset radius $R_{\rm o}$), as computed from numerical integration of (\ref{lxri}) for the selected, labelled values of the mixing exponent $m$ and shock number exponent $\q$. 
As seen from (\ref{radef}), plotting  vs. $R_{\rm a}$ can be viewed as a proxy for plotting vs. the wind density parameter $\Mdot/\vinf$.

The left vertical line at $R_{\rm a}/R_{\rm o}=1$ represents the transition from low-density  adiabatic shocks, for which $L_{\rm x} \sim \Mdot^2$,  to intermediate-density radiative shocks, for which $L_{\rm x} \sim \Mdot^{1-m}$.

The right vertical line represents  the transition from optically thin to thick winds, implemented here through a sudden change in the integration lower bound, $R_{\rm i} = \max (R_1,R_{\rm o})$, with the shock onset radius fixed at  $R_{\rm o} = 1.5 \Rstar$, and $R_1$ the radius for unit optical depth.
Like $R_{\rm a}$,  $R_1$ scales with $\Mdot/\vinf$, with values that are a small, fixed fraction $R_1/R_{\rm a} \equiv  f_{1a} $ of the adiabatic radius.
Since observed X-rays are typically a modest factor higher energy than the characteristic wind shock (e.g., $E_{\rm kev} \approx 1 \approx 2 T_{\rm kev}$),
the curves plotted in  figure \ref{lxvsrab1} assume, following (\ref{r1def}) and (\ref{radef}), $ f_{\rm 1a} = R_{\rm 1}/R_{\rm a} =  \kappa_{\rm bf}/\kappa_{\rm c} \approx 1/100$.
For very large $R_{\rm a} > R_{\rm o}/f_{\rm 1a} \approx 100 R_{\rm o}$, we thus have $R_1 > R_{\rm o}$, leading to a declining $L_{\rm x}$, as predicted by the  optically thick wind scaling $L_{\rm x} \sim R_{\rm a}^{1-\q}$ from (\ref{lxmd3}).

But for moderately dense winds,  with $R_{\rm o} < R_{\rm a} < R_{\rm o}/f_{\rm 1s}$ (between the vertical lines in the figure), the increasing $L_{\rm x}$  approaches the power-law variation $R_{\rm a}^{1-m}$ predicted by (\ref{lxmd2}).
In particular, the black curve with $\q=2$ and $m=0.4$ represents the preferred model with sub-linear scaling in $R_{\rm a}$, and thus in $\Mdot/\vinf$, implying a nearly linear scaling of $L_{\rm x}$ with $L_{\rm bol}$.

Specifically, for typical values for stellar radius ($\Rstar \approx 10-20 \Rsun$) and wind terminal speed ($\vinf \approx 2000$~km~s$^{-1}$), this intermediate-density regime with $L_{\rm x} \sim L_{\rm bol}$  applies to wind mass loss rates that range from below $10^{-7} \Msun \, {\rm yr}^{-1}$ to a few times $10^{-6} \Msun \, {\rm yr}^{-1}$;
this essentially encompasses the entire O-star spectral range for which the $L_{\rm x} \sim 10^{-7} L_{\rm bol}$ relation is found to hold.

\section{Concluding Summary}
\label{sec:summary}

The central result of this paper is that, in the common case of moderately dense O-star winds with radiative shocks  ($R_{\rm a} > R_{\rm o}$), thin-shell mixing can lead to this {\em sub-linear} scaling 
of the X-ray luminosity with the mass-loss rate, 
$L_{\rm x} \sim (\Mdot/\vinf)^{1-m}$.
Depending on the secondary scalings of wind density with bolometric luminosity, one finds that $ m \approx 0.2 - 0.4$ can  give roughly the {\em linear} $L_{\rm x}$-$L_{\rm bol}$ law that is empirically observed for O-star X-rays.
Further simulation work will be needed to see if such mixing exponent values are appropriate, and indeed to test the validity of the basic mixing exponent {\it ansatz}.

But in the course of exploring this idea of thin shell mixing, the analysis here has lead to several interesting secondary results with validity and implications that are largely independent of mixing or any specific model for it. A summary list includes:

\begin{enumerate}
\item{}
In contrast to previous analyses that invoked a density-squared emission measure description for shock production of X-rays, we derive here a more general bridging law showing how the density-squared scaling of adiabatic shocks transitions to a single-density scaling for radiative shocks. 

\item{}
For radiative shocks, the X-ray volume filling factor is not fixed (as is commonly assumed), but is reduced by the narrow extent of the shock layer, $f_{\rm v} \sim \ell/r$.

\item{}
For nearly all O-stars, the large radiative-adiabatic transition radius, $R_{\rm a} \gg R_{\rm o} \approx 1.5 \Rstar$, implies that instability-generated shocks in the wind acceleration region should follow the radiative scaling, giving the X-ray luminosity a linear scaling with mass loss rate, $L_{\rm x} \sim \Mdot/\vinf$.

\item{}
For low-density winds of lower luminosity (early B) stars, shocks should indeed become adiabatic,  implying then a steep ($L_{\rm x} \sim \Mdot^2$) decline of X-ray luminosity, as is in fact generally found for single, non-magnetic, early B-type stars, 
for which the inferred X-ray emission measure often approaches that of the full wind
 \citep{Cohen97, Cohen08b}

\item{}
Matching observed X-ray emission lines with such models of radiative shocks with or without thin-shell mixing requires the shock number to have a moderately steep decline above the X-ray onset radius, with power-law exponent $\q \approx 1.5- 2$.

\item{}
This in turn implies that the scaling of X-ray luminosity for dense, optically thick winds should saturate and even decline with increasing mass loss rate.

\end{enumerate}

This last result on X-ray absorption is perhaps not too relevant for most of the O-stars following the $L_{\rm x} - L_{\rm bol}$ relation, for which optical depth effects are weak to marginal.
But it can become important for the dense, moderately optically thick winds of extreme, early O-stars like HD93129A,  which can be viewed as a transitional object to the WNH type Wolf-Rayet stars.
More generally, the high-density of Wolf-Rayet winds imply that absorption should strongly attenuate X-rays from any instability-generated shocks in their wind acceleration region.
As such, the observed hard X-rays seen from Wolf-Rayet stars like WR6 (EZ CMa) seem unlikely to be explained by this standard model of  LDI shocks \citep{Oskinova12}.
This also has potential implications for interpreting observed X-rays from very massive stars that have $L_{\rm x} \sim 10^{-7} L_{\rm bol}$ despite having very high wind optical depths \citep{Crowther10}, and whether these might instead originate from  wind-wind collisions of close, undetected binary companions.

Indeed, the mixing {\em ansatz} in this paper could also be applied to model X-ray emission from colliding wind binaries, and their $L_{\rm x}$ scaling with orbital separation. Wide binaries with adiabatic shocks should still follow the usual inverse distance scaling, as directly confirmed by observations of multi-year-period elliptical systems like WR140 and $\eta$~Carinae \citep{Corcoran11}. But in close, short (day to week) period binaries with radiative shocks \citep{Antokhin04}, mixing could reduce and limit the effective X-ray emission from the wind collision \citep{Parkin10}, and thus help 
clarify
why such systems often hardly exceed the $L_{\rm x} \approx 10^{-7} L_{\rm bol}$ scaling found for 
single stars 
\citep{Oskinova05,Corcoran11,Gagne11}.

Finally, in addition to exploring such effects in colliding wind binaries,
a top priority for future work should be to carry out detailed simulations of the general effect of thin-shell mixing on X-ray emission, and specifically to examine the validity of this
 mixing-exponent {\em ansatz} for modeling the resulting scalings for X-ray luminosity.

\section*{Acknowledgments}
This work was supported in part by NASA ATP grant
NNX11AC40G to the University of Delaware. 
D.H.C. acknowledges support from NASA
ADAP grant NNX11AD26G and NASA {\it Chandra} grant AR2-13001A to Swarthmore College.
 
\bibliographystyle{mn2e}
\bibliography{OwockiS}

\end{document}